\documentclass{aa}
\usepackage{graphicx}
\usepackage[varg]{txfonts}
\def\lsim{\!\!\!\phantom{\le}\smash{\buildrel{}\over
 {\lower2.5dd\hbox{$\buildrel{\lower2dd\hbox{$\displaystyle<$}}\over
                                 \sim$}}}\,\,}
\def\gsim{\!\!\!\phantom{\ge}\smash{\buildrel{}\over
{\lower2.5dd\hbox{$\buildrel{\lower2dd\hbox{$\displaystyle>$}}\over
                               \sim$}}}\,\,} 
\def\asec{\ifmmode ^{\prime\prime}\else$^{\prime\prime}$\fi}

\begin{document}

\title{Sub-arcsecond-resolution LOFAR observations of bright sub-millimetre galaxies in the North Ecliptic Pole field}
\titlerunning{Sub-arcsecond resolution LOFAR observations of SMGs}
\author{M. Bondi\inst{1}
          \and I. Prandoni\inst{1}
          \and M. Magliocchetti\inst{2}
          \and L. Bisigello\inst{1,3,4}
          \and M. Bonato\inst{1}
          \and M. Giulietti\inst{1}
        \and R. Scaramella\inst{5,6}
          \and G. Brunetti\inst{1}
          \and F. Vitello\inst{7}
        }
        \authorrunning{M. Bondi et al.}        

\institute{INAF - Istituto di Radioastronomia, Via Gobetti 101, 40129, Bologna
\and INAF - Istituto di Astrofisica e Scienza dello Spazio di Roma, Via Fosso del Cavaliere 100, 00133 Roma, Italy
\and Dipartimento di Fisica e Astronomia ``G. Galilei'', Universit\'a di Padova, Via Marzolo 8, 35131 Padova, Italy
\and INAF - Osservatorio Astronomico di Padova, Vicolo dell'Osservatorio 5, 35122 Padova, Italy
\and INFN - Sezione di Roma, Piazzale Aldo Moro, 2 - c/o Dipartimento di
Fisica, Edificio G. Marconi, I-00185 Roma, Italy
\and INAF - Osservatorio Astronomico di Roma, Via Frascati 33, I-00078
Monteporzio Catone, Italy
\and INAF - Osservatorio Astrofisico di Catania, Via Santa Sofia 78, 95123 Catania, Italy
}

\abstract 
{Bright sub-millimetre galaxies (SMGs) 
contribute significantly to the star formation rate density
(20-50\%) and stellar mass density ($\sim$ 30-50\%) at redshifts 2-4 with star formation rates (SFRs) $\protect\gsim 1000$ M$_\odot$,yr$^{-1}$ and stellar masses of $\sim 10^{11}$-$10^{12}$ M$_\odot$. The number of bright SMGs with such high SFRs is hard to reconcile with the standard models of galaxy formation and evolution. In this paper we provide evidence that, in a small sample of 12 bright SMGs, the SFRs derived from spectral energy distribution (SED) fitting are significantly higher than those obtained using 
low-frequency radio emission as a proxy for star formation. Using the International LOFAR Telescope (ILT), which allows imaging
at 144 MHz with sub-arcsecond  angular resolution, we have produced
deep images of a small sample of bright SMGs in the North Ecliptic Pole (NEP) field extracted from the NEPSC2 survey. For all 12 SMGs, we find radio-emitting mid-infrared galaxies at distances from a few arcseconds down to sub-arcsecond scales from the SMG and/or the presence of a radio-emitting active galactic nucleus (AGN). The SFRs derived from the radio emission of the SMG, disentangled from the AGN-related radio emission, are systematically lower by a factor of $\sim 5$ (median value) than those derived from the multi-band SED fitting. We discuss whether our assumptions might be,
at least in part, responsible for the observed discrepancy. We argue that the radio-derived SFRs are not systematically underestimated but can be affected by a significant dispersion ($0.3-0.5$ dex). 
Considering these new SFR estimates, the offset of the specific SFR of the 12 bright SMGs from the star-forming galaxy main sequence ($\Delta\mathrm{(SSFR)}$) is significantly reduced. Using sub-arcsecond  radio images to disentangle the contribution of the AGN and the radio emission as a proxy for the determination of the SFRs, we find that all 12 bright SMGs are found in star-forming galaxies (SFGs) or hybrid SFG--AGN systems that, on average, are only a factor of 2 more star-forming than the main sequence galaxies. 
}

\maketitle

\section{Introduction}
Sub-millimetre galaxies \citep[SMGs;][]{1997ApJ...490L...5S,1998Natur.394..248B,1998Natur.394..241H,1999ApJ...515..518E}
are distant, ultra-luminous galaxies where star formation
is heavily obscured by dust. In the past 25 years, a number of SMGs have been studied, providing valuable insights into their properties \citep[for reviews see][]{2002PhR...369..111B,2014PhR...541...45C}.
Most of the SMGs have been found at $z\sim 1.5 - 3.5$ \citep{ 2000AJ....119.2092B,2005ApJ...622..772C, 2006MNRAS.370.1185P,  2007MNRAS.379.1571A, 2011MNRAS415.1479W, 2012MNRAS.420..957Y, 2012MNRAS.426.1845M, 2014ApJ...788..125S, 2015A&A...577A..29M, 2016ApJ...820...82C, 2017ApJ...840...78D, 2017A&A...608A..15B, 2017MNRAS.469..492M,2017ApJ...839...58S, 2022MNRAS.514.2915S}.
SMGs are gas-rich \citep{2005MNRAS.359.1165G, 2006ApJ...640..228T, 2008ApJ...680..246T, 2010ApJ...724..233E, 2011ApJ...739L..31R, 2011MNRAS.412.1913I, 2013MNRAS.429.3047B, 2017MNRAS.467.1222H} and massive objects \citep{
2004ApJ...617...64S,  2008MNRAS.386.1107D, 2011ApJ...740...96H, 
2012A&A...541A..85M, 2013MNRAS.432.2012T, 2015ApJ...806..110D, 
2017MNRAS.469..492M, 2022MNRAS.514.2915S}, with high star formation rates (SFRs) ranging from 100 to $>1000$ M$_\odot$\,yr$^{-1}$ \citep{2010A&A...514A..67M, 2012ApJ...761...89B, 2014ApJ...784....9B, 2022MNRAS.514.2915S}, which places them on, or above, the massive end of the star-forming main sequence (MS). The impact of an active galactic nucleus (AGN) on the observed properties of SMGs is still unclear.
Even though the AGN component rarely dominates the sub-millimetre emission, it is not uncommon for SMGs to host AGNs \citep{2003AJ....125..383A, 2005ApJ...632..736A,  2010MNRAS.401.2763L, 2013ApJ...778..179W,  2018ApJ...853...24U}.
Finally, the spatial clustering and environment properties of SMGs show that they
reside in high-mass dark matter haloes \citep{2004ApJ...611..725B, 
2007MNRAS.375.1121M, 2016ApJ...831...91C, 2017MNRAS.464.1380W}.
As SMGs are significant contributors to the star formation history out to a redshift of at least 5 and
 potential precursors of the present-day massive elliptical galaxies, understanding the physical nature of SMGs is crucial for constructing a self-consistent galaxy evolution model \citep[e.g.][]{2006MNRAS.371..465S, 
 2013Natur.498..338F,2014ApJ...782...68T, 2014ApJ...788..125S}. 
In particular, the number density of the brightest SMGs with SFR$\gsim 1000$ M$_\odot\,$yr$^{-1}$  is several times higher than that derived from model predictions. Neither mergers nor cold accretion could produce SFR values $\gsim 1000$ M$_\odot$\,yr$^{-1}$ \citep{2005MNRAS.363....2K, 
2010MNRAS.404.1355D, 2010MNRAS.407.1701N, 2015Natur.525..496N}, in contrast with the values derived for the brightest SMGs.

This tension has been qualitatively explained by arguing that the photometric redshifts and SFRs derived from spectral energy distribution (SED) fitting may be
strongly affected by dust and therefore overestimated. Furthermore, a possible contribution to the infrared emission of an AGN that is not fully taken into account by the fitting codes might undermine the derived parameters \citep[e.g.][]{2017MNRAS.465.1401S}.
In this paper we present and discuss images of SMGs obtained with the International LOw-Frequency ARray \citep[LOFAR;][]{2013A&A...556A...2V} Telescope (ILT). The ILT exploits the contribution of the 14 international stations (ISs), which provide baselines up to $\sim 2000$ km  and therefore enable imaging at
144 MHz with sub-arcsecond angular resolution.
We used the ILT to image, with unprecedented resolution and sensitivity, a well-defined small sample of bright SMGs selected at 850\,$\mu$m in the North Ecliptic Pole (NEP) field. 

In Sect.~\ref{sec:smg} we introduce the sample of 12 bright SMGs derived from the larger sample published by \citet{2022MNRAS.514.2915S}. Section~\ref{sec:obs} describes the ILT observations and data analysis, while Sect.~\ref{sec:res} contains the images obtained and, for each source, a brief description of the results. The discussion and interpretation of our findings is the subject of Sect.~\ref{sec:dis}.
Throughout this paper we assume $\Omega_M=0.3$, $\Omega_\Lambda=0.7$ and 
$H_0=70$ km\,s$^{-1}$\,Mpc$^{-1}$.

\begin{table*}
\caption{Bright SMGs in the NEP region.}
\centering
\begin{tabular}{rccrccccl}
\hline\hline
SMM ID&RA&Dec.&S$_{850\mu m}$&redshift&LIR&SFR&Mstar&Class \\
      &  (deg)  &  (deg)  & (mJy)  &        & ($\times 10^{39}$ W\,Hz$^{-1}$)& (M$_\odot$\,yr$^{-1}$)& ($\times 10^{10}$ M$_\odot$) & \\ 
3  & 268.1823 & 66.143 & $23.2\pm 1.9$ & $2.89\pm 0.15$ & $8.82\pm 1.16$ & $1965\pm 309$ & $\,\,\,82.7\pm 15.8$ & SFG \\
12 & 269.3376 & 65.928& $11.7\pm 3.2$ & $2.59\pm 0.13$ & $7.58\pm 0.94$ & $2276\pm 312$ &$\,\,\,29.7\pm \,\,\,4.4$ & SFG \\
14 & 269.5642 &65.867& $10.1\pm 4.0$ & $1.73\pm 0.09$ &$2.57\pm 0.71$ & $437\pm 132$ & $131.1\pm 31.2$ & AGN--SFG \\
15 & 267.9203 & 66.803 & $12.1\pm 2.5$ & $2.10\pm 0.80$ & $2.73\pm 2.21$ & $932\pm 796$ & $\,\,\,\,\,\,4.4\pm \,\,\,1.8$ & SFG \\
16 & 268.1928 & 66.103 & $10.9\pm 2.9$ &$3.21\pm 0.07$ & $9.35\pm 0.83$ & $2031\pm 235$ &$171.7\pm 17.9$ & AGN--SFG \\
24 & 270.4636 & 66.574 & $9.8\pm 2.8$ & $2.12\pm 0.16$ & $9.20\pm 1.76$ & $3137\pm 590$ & $\,\,\,38.5\pm \,\,\,7.3$ & SFG \\
29 & 268.8129 & 66.732 & $11.0\pm 1.8$ & $3.38\pm 0.19$ & $1.98\pm,0.54$ & $544\pm 144$ & $\,\,\,\,\,\,5.8\pm \,\,\,1.5$ & AGN\\
47 & 268.3128& 66.830& $10.4\pm 1.7$ & $3.26\pm ,0.63$ &$8.21\pm 2.88$ & $2037\pm 714$ &$\,\,\,40.1\pm 33.9$ & AGN--SFG \\
49 & 268.3026 &66.980 & $9.7\pm 2.2$ & $2.57\pm 0.20$ & $6.33\pm 1.74$ & $1883\pm 522$& $\,\,\,17.1\pm \,\,\,5.1$ & SFG \\
55 & 268.6999 & 66.580 & $10.0\pm 1.8$ & $3.13\pm 0.10$ & $7.20\pm 0.69$ &$1386\pm 133$ & $ 154.3\pm 14.8$ & AGN--SFG \\
74 & 268.3215& 66.849& $9.1\pm 1.7$ &$2.99\pm 1.01$ & $5.67\pm 2.92$ & $1626\pm 870$ & $\,\,\,27.7\pm 33.9$ & AGN--SFG \\
77 & 268.2363 & 66.723& $9.0\pm 1.8$ & $3.57\pm 1.13$ & $5.52\pm 3.06$ & $1428\pm 814$ &$\,\,\,26.0\pm 23.3$ & AGN--SFG \\
\hline
\end{tabular}
\tablefoot{Col. 1: SMM ID name. Cols. 2 and 3: Right ascension and declination of the IRAC counterpart in degree. Col. 4: Sub-millimetre flux density from SCUBA-2 at 850\,$\mu$m.  Col.5: Photometric redshift. Col.6: Infrared luminosity in units of $10^{39}$ W\,Hz$^{-1}$. Col. 7: Star formation rate in M$_\odot$\,yr$^{-1}$. Col. 8: Stellar mass in units of 10$^{10}\,\mathrm{M}_\odot$.
Col. 9: Classification as AGN, SFG, or hybrid AGN--SFG (see Sect. \ref{sec:AGN}), based on the AGN fraction parameter derived by SED fitting in \citet{2022MNRAS.514.2915S}.}
\label{tab:submm}
\end{table*}

\section{Sub-millimetre galaxies in the NEP region}
\label{sec:smg}
The NEP region has been targeted by the \textit{James Clerk Maxwell} Telescope (JCMT) Submillimetre Common User Array-2 
\citep[SCUBA;][]{2013MNRAS.430.2513H} as one of the fields of the Cosmology Legacy Survey \citep[S2CLS,][]{2017MNRAS.465.1789G}. The original area covered by the S2CLS in the NEP region at 850 $\mu$m was 0.6 deg$^2$ with an rms sensitivity of 1.2 mJy\,beam$^{-1}$. This yielded a catalogue of 330 sources with S/N $>3.5$ \citep{2017MNRAS.465.1789G}. More recently, the North Ecliptic Pole Scuba-2 Survey (NEPSC2) expanded the area imaged at 850 $\mu$m to 2 deg$^2$ with a rms sensitivity in the range 1-2 mJy\,beam$^{-1}$ \citep{2020MNRAS.498.5065S}. The NEPSC2 catalogue contains 549 sources with S/N $> 4$ at $850~\mu$m.
The number of unique sources obtained by merging the two catalogues
is 647 and the process of multi-wavelength counterpart identification was
performed by \citet{2022MNRAS.514.2915S} using ancillary datasets ranging from the optical to the radio band. 
We refer to \citet{2022MNRAS.514.2915S} for the details of the optical and near-infrared (NIR) counterpart identification process and the assembly of the multi-wavelength
catalogue that contains homogeneous photometry, as well as physical properties obtained through SED fitting.
We recall that 449 SMGs are reliably identified
and that a subsample  of 272 sources (hereafter referred to as the clean sample)
was obtained after removing all SMGs with possible multiple
identifications and keeping only sources with good-quality best-fitting SED.

A study of the radio properties of the full `clean' sample is deferred to
a different paper in preparation. Here we focus on the brightest
SMGs in the `clean' sample by selecting all the sources
with $S_\mathrm{850\mu m} > 9$ mJy. Furthermore, we only  considered the SMGs within 1.2 deg of the field centre of the LOFAR observations. This is not only the most sensitive region but also the one that allows proper imaging and analysis using the ISs, essential to obtain sub-arcsecond resolution at 144 MHz.
With these constraints we obtained a sample of 12 SMGs, which are listed
in Table~\ref{tab:submm} with the Infrared Array Camera (IRAC) positions and some relevant physical parameters listed from the catalogue
published in \citet{2022MNRAS.514.2915S}. 

\section{LOFAR observations and data analysis}
\label{sec:obs}

The NEP region is one of the field of the LOFAR Two-Meter Sky Survey (LoTSS) Deep Fields project \citep{2023MNRAS.523.1729B}. This field was chosen because the NEP is the location of the Euclid Deep Field North (EDFN), one of the deep fields observed by the Euclid mission \citep{2022A&A...662A.112E}, and the 
only one in the northern sky. Euclid observations will provide sub-arcsecond optical and NIR imaging down to $H=26$ mag over a 20 deg$^2$ field centred at RA$=269.73$ deg and Dec.$=+66.02$ deg. 
The NEP field was observed for 72 hours with LOFAR during cycle 12 (proposal LC12\_027, P.I. van Weeren). The observations were distributed over nine nights in the period June to November 2019.
\citet{2024A&A...683A.179B} presents a $6\asec$ resolution image obtained combining all observations, but excluding the ISs, as well as the catalogue of the radio sources with signal-to noise ratio S/N$>5$, extracted in the inner 10
square degrees circular region centred on the EDFN coordinates.
Eleven out of the 12 sources listed in Table~\ref{tab:submm} have a radio counterpart in the $5\sigma$ catalogue.
The remaining source, SMM ID 74\footnote{To identify SMGs we use the 850-$\mu$m source ID of the combined catalogue published by \citet{2022MNRAS.514.2915S}}, has a $4.3\sigma$ radio counterpart, i.e. below the catalogue threshold.

In this paper we present sub-arcsecond resolution ILT 
images at 144 MHz obtained combining the ISs.
Only the six nights with the best
rms noise, on the basis of the $6\asec$ resolution imaging, were chosen for this analysis.
The observations were carried out in the standard survey setup, $\sim 8$ hours
on-source placed between two short ($\sim 10$ min) observations
of a flux density calibrator, in this case
3C295 and 3C48. A total of 51 stations (24 cores, 14 remote, and 13 international)\footnote{At the epoch of the observations, the number of ISs was 13.} were used for these observations, with the exception of one night that only had 12 ISs.

Each dataset was calibrated against direction-independent errors using \texttt{\small PREFACTOR} \citep{2019A&A...622A...5D, 2016ApJS..223....2V,
2016MNRAS.460.2385W}\footnote{https://github.com/lofar-astron/prefactor}
with the calibrator 3C48. A high-resolution model appropriate to  calibrate the IS was adopted.
Then the datasets went through \texttt{\small DDF-pipeline} to perform the direction-dependent calibration \citep{2019A&A...622A...1S, 2021A&A...648A...1T}\footnote{https://github.com/mhardcastle/ddf-pipeline}.
Finally, the calibration of the IS data was performed using the LOFAR long baseline pipeline (version 3.0). The original pipeline is described in detail in  \citet{2022A&A...658A...1M}, and here we only provide a brief summary of the steps we performed for each of the six nights.

Firstly, the calibration solutions from \texttt{\small PREFACTOR} and \texttt{\small DDF-pipeline} were applied to the un-averaged data, and the data were sorted by frequency and split into
 25 datasets, each with 2 MHz bandwidth, covering the frequency range 121 MHz - 168 MHz. At this stage, the time and frequency resolution are 1s and
 12.25\, kHz, respectively.
The datasets were then phase-shifted to the position of the delay calibrator source, primary-beam corrected and averaged in time (8s) and frequency (98 kHz). The core stations  were combined together to form a super-station in order to increase the sensitivity to the long baselines and all the 25 2 MHz bandwidth files were concatenated into a single dataset.
The next step was the calibration of the ISs. The dataset of the delay calibrator was processed using the script
 {\tt facetselfcal.py}\footnote{https://github.com/rvweeren/lofar\_facet\_selfcal} developed by \citet{2021A&A...651A.115V} to iteratively 
 derive the delay factors to be applied to the ISs and to self-calibrate the phases and amplitudes of the whole array.
 Finally,  phase-shifted, beam corrected and averaged datasets at the position of the 12 sources to be imaged (hereafter target sources) were generated, 
 and the phase and amplitude solutions obtained from the delay calibrator
 were applied to the data of the target sources. 
 We note that the {\tt facetselfcal.py} script has recently been incorporated into
 a new version of the LOFAR long baseline pipeline, and all the above steps can be performed automatically.

 \begin{figure}[htp]
 \centering
 \includegraphics[width=8cm]{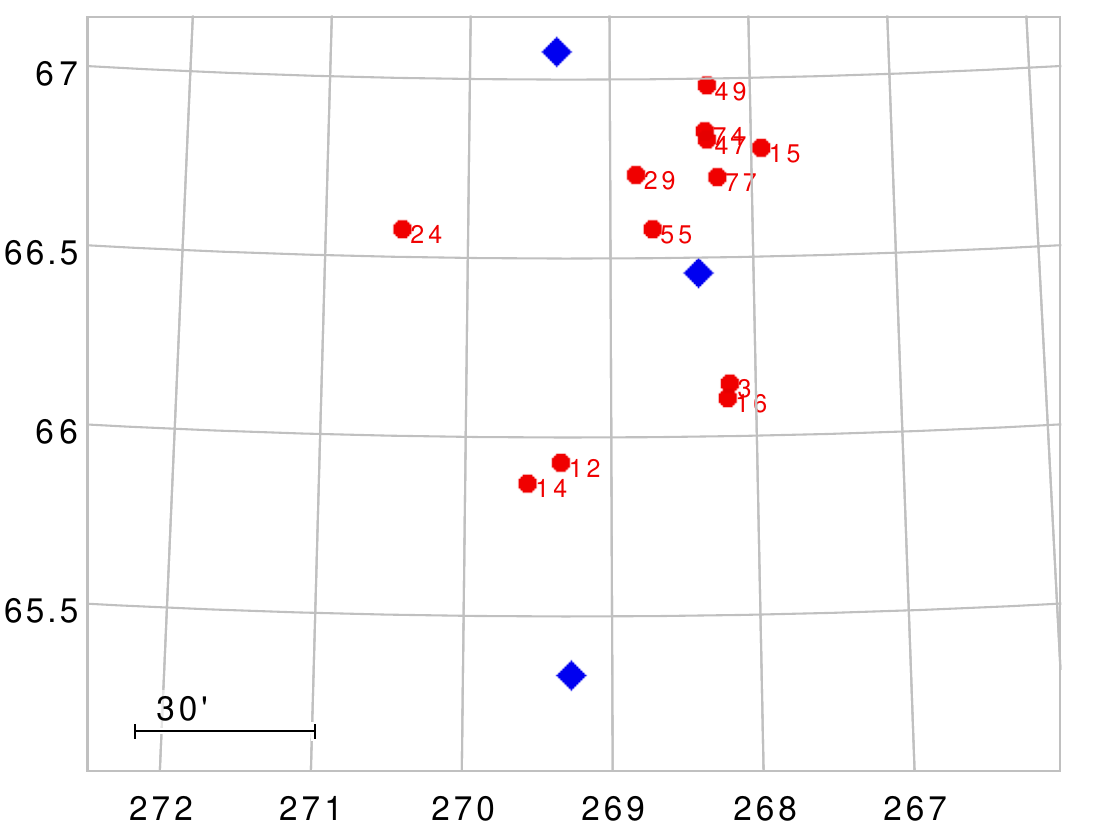}
 \caption{Sky plot showing the RA and Dec. positions of the 12 SMGs (red points) and the three bright calibrator sources that were used to provide phase and gain corrections to refine the calibration of the SMGs before imaging.}
 \label{fig:skypos}
\end{figure}

All 12 SMGs listed in Table~\ref{tab:submm} are too faint
 (peak flux densities $\lsim 1$ mJy\,beam$^{-1}$ at the resolution of $6\asec$) to be directly self-calibrated. Therefore, 
 to refine the calibration towards our SMGs
 we used  bright ($\gsim 100$ mJy\,beam$^{-1}$) sources closer to our target sources than the delay calibrator. In Fig.\ref{fig:skypos} we show the sky positions of our 12 bright SMGs (red points) and of the 3 radio bright sources whose phase and gain solutions were applied to the target sources before imaging (blue diamonds). For each target source, we applied the solutions from the closest of the three bright sources. When two bright sources
 were at a comparable distance to a target source, we processed the target twice, using the solutions derived for each bright source, and compared the results. In all these cases we find consistent results 
 in terms of source structure and fluxes, and minor differences ($\lsim 10\%$) in the rms of the final images. In these cases we decided to apply the calibration that produced the images with the smallest rms.
 Finally, for each target source, we imaged together the six fully calibrated datasets using WSCLEAN \citep{2014MNRAS.444..606O}. By varying the weighting scheme and the {\it uv} taper, for each source we produced images
 at different angular resolutions: a full resolution image with a full width at half maximum (FWHM) of  $\simeq 0.3\asec$, an intermediate resolution image with FWHM $\simeq 0.5\asec$
 and a low resolution image with FWHM$\simeq 1\asec$. 
 Combining together six nights, we obtain 144 MHz ILT images at sub-arcsecond resolutions with an rms noise down to 13 $\mu$Jy\,beam$^{-1}$.
 
\section{Results}
\label{sec:res}

\begin{figure}[htp]
 \centering
 \includegraphics[width=5.3cm]{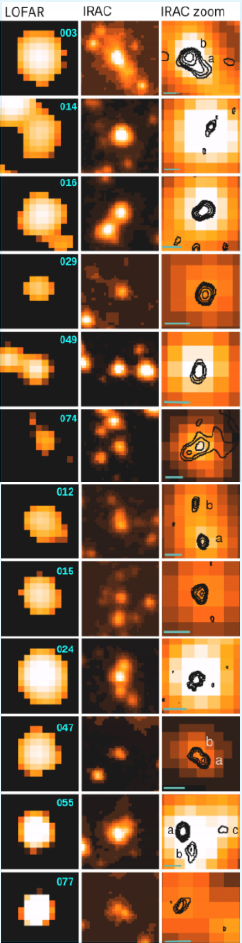}
 \caption{Images of the 12 bright SMGs. For each galaxy (the SMM ID is displayed in the top-right corner of the left panel) the left and middle panels display the LOFAR $6\asec$ and the IRAC $4.5\mu$m image, respectively. The two images show the same $20\asec\times 20\asec$ region centred at the position of the SMG. The right panel is a zoomed-in sub-region of the IRAC image with the 144 MHz ILT radio contours drawn in black. The sizes of the zoomed-in sub-regions are in the range $2.5\asec - 6\asec$, and the length of the cyan ruler in the lower-left corner of the IRAC zoomed-in images is $1\asec$. The letters (a, b, and c) refer to the different galaxies or components listed in 
 Table~\ref{tab:radio}. Radio contours are drawn at $3, 4, 5, 7$, and $10\times \sigma$, where $\sigma$ is the local radio rms noise.}
 \label{fig:images}
\end{figure}

For each of the 12 SMGs, in Fig.~\ref{fig:images} there are three different images (from left to right): the $20\asec \times 20\asec$ region centred at the galaxy position in the $6\asec$ resolution LOFAR image obtained from the data reduction of the full dataset \citep{2024A&A...683A.179B} without the use of the ISs and labelled with the source SMM ID; the same region in the IRAC $4.5\mu$m mid-infrared (MIR) image obtained from the Cosmic Dawn Survey mosaic of the EDFN published by  \citet{2022A&A...658A.126E}; a zoomed-in region of the IRAC $4.5\mu$m image with overplotted the radio contours from the ILT image. 
The radio properties of the SMGs derived from the ILT images are listed in Table~\ref{tab:radio}.
The LOFAR images (at $6\asec$ and sub-arcsecond resolution) are used to identify MIR galaxies emitting at 144 MHz within $\sim 10\asec$ of the SMGs. Such objects could contaminate the far-infrared (FIR) and sub-millimetre fluxes used in the SED fitting of the SMG, and therefore yield to an overestimate of the SFR values.

\begin{table*}
\caption{Fitted radio properties of bright SMGs.}
\centering
\begin{tabular}{rcccrccccrcc}
\hline\hline
SMM ID& RA & Dec. & FWHM & P.A.& rms & $S_p$ & $S_T$ & Size & PA &Multi&$\log{T_{\rm b}}$\\
      &  (deg)&  (deg)&(arcsec$^2$)&(deg) &($\mu$Jy/bm)&($\mu$Jy/bm)&($\mu$Jy)&(arcsec$^2$)&(deg)&  &(K)\\
3a &268.18213&66.1430&$1.07\times 0.86$&109& 26 & 155&~~467&$1.56\times 1.14$& 34& Y&4.78\\
3b &268.18280&66.1432&                 &   &    & 303&~~421&$0.75\times 0.42$ & 153& &5.48   \\
12a&269.33762&65.9276&$0.54\times 0.39$&174& 13 & 112&~~180&$0.45\times 0.24$& 27& Y&5.55\\
12b&269.33786&65.9282&                 &   &    &~~72&~~121&
$0.80 \times\sim 0.1$ &  2& & 5.50   \\
14 &269.56387&65.8668&$0.54\times 0.39$&173& 13 &~~75&~~176&$1.11\times \sim0.1$ & 159& Y&5.39\\
15 &267.92029&66.8030&$0.38\times 0.26$&176& 20 & 209&~~549&$0.47\times 0.35$&  5& N&5.78\\
16 &268.19272&66.1036&$0.56\times 0.41$&172& 13 &214&~~479&$0.69\times 0.34$&109& N&5.70\\
24 &270.46381&66.5738&$0.27\times 0.19$&179& 18 &305&1152&$0.40\times 0.33$ &136& N&6.20\\
29 &268.81271&66.7325&$0.58\times 0.43$&172& 15 &176&~~234&$0.31\times 0.23$&102& N&5.92\\
47a&268.31247&66.8295&$0.63\times 0.47$&168& 17 &193&~~285&$0.38\times 0.34$& 84& Y&5.75\\
47b&268.31287&66.8297&                 &   &    &150&~~263&$0.55\times 0.38$&121&  &5.49  \\
49 &268.30255&66.9804&$0.62\times 0.47$&168& 19 &125&~~192&$0.50\times 0.31$&157& Y&5.41\\
55a&268.70044&66.5806&$0.57\times 0.42$&172& 15 &231&~~323&$0.37\times 0.26$&175& Y&5.91\\
55b&268.70010&66.5802&                 &   &    &~~73&~~264&$0.99\times 0.63$&  2& & 5.00   \\
55c&268.69882&66.5806&                 &   &    &~~57&~~134&$0.83\times 0.21$& 89& & 5.26  \\
74 &268.32144&66.8491&$3.11\times 1.66$&111&22 &~~96&~~156&$3.03\times 1.10$& 130& Y&4.05\\ 
77 &268.23682&66.7231&$0.37\times 0.26$&176&18 &105&~~390&$0.69\times 0.34$& 134& N&5.65\\
\hline
\end{tabular}
\tablefoot{Col. 1: Source ID. Cols. 2 and 3: Right ascension and declination measured from  the ILT images. Cols 4 and 5: Restored beam (FWHM major and minor axis and position angle) of the image used to derive the radio parameters. Col. 6: rms noise. Cols. 7 and 8: Peak brightness and total flux of the fitted Gaussian component. Col .9 and 10:
Fitted deconvolved angular size (FWHM major and minor axis \& position angle). Col. 11: Multiplicity flag. Col. 12:
$\log$ value of the brightness temperature.
}
\label{tab:radio}
\end{table*}
    
Below we give, for each SMG, a brief description  of its radio properties.

\paragraph{SMM ID 3.} It is the brightest SMG in our sample with $S_{850\mu m}=23.2$ mJy at a redshift $z=2.89$. The radio contours are from a tapered ILT image with an angular resolution of $1.07\asec \times 0.86\asec$. The rms noise in this tapered image is $26\,\mu$Jy\,beam$^{-1}$. The radio source is fit with two components listed as 3a and 3b in Table~\ref{tab:radio}. The MIR counterpart of the SMG is closer to the southern radio component 3a. The total flux of the two components detected at the $\sim 1\asec$
resolution is consistent with that measured at $6\asec$; therefore, no significant extended radio flux is missing in the $\sim 1\asec$ resolution radio image. Component 3a is weaker (S/N$\simeq 6.0$), extended and is almost resolved out in images with better angular resolution ($\sim 0.3\asec$) leaving only a patchy low brightness (S/N$\lsim 3$) radio emission, while radio component 3b is brighter (S/N$\simeq 11$) and more compact.
The brightest (white) region in the $4.5\mu$m image shown in Fig.~\ref{fig:images}  can be fitted with two components, whose
positions are consistent with those detected at 144 MHz supporting the scenario with two
different MIR- and radio-emitting objects separated by $1.3\asec$, corresponding to $\sim 10$ Kpc at $z=2.89$, assuming the two objects are at the same redshift. We note that other fainter IRAC sources are located within a few arcsecond of the SMG, but none of these has a radio counterpart. Indeed, the high sub-millimetre flux suggests this could be a starburst-dominated major merger between two gas-rich galaxies, such as that in HS1700.850.1 \citep{2023MNRAS.523.2818P}. Higher resolution NIR imaging (as those provided by the Euclid mission) or molecular spectral line observations are needed to test this scenario.
\paragraph{SMM ID 12.} The ILT radio image has a resolution of $0.54\asec \times 0.39\asec$ and an rms noise of $13\,\mu$Jy\,beam$^{-1}$. Two radio components are clearly detected in this ILT image, and 
are centred on two galaxies visible in the zoomed-in IRAC $4.5\mu$m image. We fitted the IRAC emission with two Gaussian components and find that the IRAC galaxies are coincident
in position with the radio sources and 
separated by $\sim 2.2\asec$ corresponding to $\sim 18$ kpc at $z=2.59$ (assuming that the two galaxies are at the same redshift in a merging system). 
The southern component, labelled 12a, is coincident with the optical/NIR counterpart of the SMG \citep{2022MNRAS.514.2915S}. Component 12a is brighter (S/N$\simeq 9$) and more compact (in both the ILT and IRAC images), while the northern one, labelled 12b, is weaker (S/N$\simeq 6$) and extended in the north--south direction. There is some faint extended emission around component 12a that is better recovered in the $\sim 1\asec$ image. By comparing the flux density measured at $6\asec$ with the sum of the flux density of 12a and 12b measured from the $\sim 1\asec$ resolution image, we find that less than $10\%$ of the radio flux is missing. 
\paragraph{SMM ID 14.} The ILT radio image has a resolution of $0.54\asec \times 0.39\asec$ and an rms noise of  $13\,\mu$Jy\,beam$^{-1}$. A single radio component, elongated in the south-eastern direction, is detected at the position of the SMG with a S/N just above 5. 
The flux density derived from this sub-arcsecond resolution image is significantly lower (a factor 2.5) than that measured at $6\asec$ resolution, and 20-30 per cent of the flux is still missing from the ILT image at $\sim 1\asec$ resolution. 
A bright radio source, coincident in position with another MIR source, is found at about $10\asec$ in the NE direction (see Fig.~\ref{fig:images}). The FIR and sub-millimetre photometry could be contaminated by this radio source. 
\paragraph{SMM ID 15.} The ILT radio image has a resolution of $0.38\asec \times 0.26\asec$ 
and an rms noise of $20\,\mu$Jy\,beam$^{-1}$. At this resolution the SMG is detected with a S/N$\simeq 10$ and the total flux is only $10\%$ lower than that measured at $6\asec$ resolution. The source is resolved in both the N--S and E--W directions. 
\paragraph{SMM ID 16.} The ILT radio image has a resolution of $0.56\asec \times 0.41\asec$ and an rms noise of $13\,\mu$Jy\,beam$^{-1}$. The source  is detected with a S/N$\simeq 16$ and it
is slightly resolved. At this resolution the total flux detected on the sub-arcsecond
scale is about half of that measured at $6\asec$ resolution. Roughly 30\%
of the flux detected at $6\asec$ is still missing in the ILT image at low
($\sim 1\asec$) resolution.
\paragraph{SMM ID 24.} The ILT radio image has a resolution of $0.27\asec \times 0.19\asec$ 
and an rms noise of $18\,\mu$Jy\,beam$^{-1}$. The radio source shows a relatively bright
core (S/N$\simeq 17$) embedded in weak and diffuse emission. Only $\sim 60\%$ of the total flux is recovered at
resolution $\lsim 1 \asec$ with respect to the flux measured at $6\asec$ resolution. 
In the MIR image the galaxy has nearby companions but none of these shows significant
radio emission with the present sensitivity.
\paragraph{SMM ID 29.} The ILT radio image has a resolution of $0.58\asec \times 0.43\asec$ 
and an rms noise of $15\,\mu$Jy\,beam$^{-1}$. A single compact radio source with a S/N$\simeq 12$ is detected. About $30\%$ ($15\%$) of the flux measured at $6\asec$ resolution is missing in this image (in the $\sim 1\asec$ resolution image).
This is the most compact radio source and the only one, of the 12 SMGs discussed in this paper, that was unambiguously classified as an AGN-dominated object by the SED fitting in \citet{2022MNRAS.514.2915S}.
\paragraph{SMM ID 47.} The ILT radio image has a resolution of $0.63\asec \times 0.47\asec$ 
and an rms noise of $17\,\mu$Jy\,beam$^{-1}$. The radio source has a double-component morphology. The MIR image is extended in the same direction as the radio one, and it can be 
fitted with two Gaussian components whose positions are
coincident with the radio ones.
The southern radio component, labelled 47a in Table~\ref{tab:radio}, has a S/N$\simeq 11$ and is associated with the MIR counterpart of the SMG. The northern component, 47b in Table~\ref{tab:radio}, has a S/N$\simeq 9$.
Some extended radio emission embedding the two components is present and is better accounted for in the $\sim 1\asec$ resolution image, which recovers a total flux within $10\%$ of that measured in the $6\asec$ resolution image.
\paragraph{SMM ID 49.} The ILT radio image has a resolution of $0.62\asec \times 0.47\asec$ 
and an rms noise of $19\,\mu$Jy\,beam$^{-1}$. At this resolution, a single radio component (S/N$\simeq 7$) accounting for only half of the flux measured at $6\asec$ resolution is detected, The $\sim 1\asec$ resolution image accounts for all the flux measured at lower resolution. Another radio source associated 
with a MIR galaxy is within $10\asec$ east of the SMG; therefore,
the FIR and sub-millimetre fluxes could be contaminated and overestimated.
\paragraph{SMM ID 55.} The ILT radio image has a resolution of $0.57\asec \times 0.42\asec$ 
and an rms noise of $15\,\mu$Jy\,beam$^{-1}$. Three radio components are detected within a
radius of $\sim 2\asec$. The western and faintest component (labelled 55c in Table~\ref{tab:radio}, S/N$\simeq 4$) is
associated with a different MIR source. At the position of the SMG, we find two components (55a and 55b in Table~\ref{tab:radio}).The northern component, 55a, is brighter and more compact with a S/N$\simeq 15$, while the southern one, 55b,  has a fainter peak brightness with S/N$\simeq 5$ and is more extended. Component 55a is closer to the MIR and optical counterpart of the SMG. At this resolution the total flux of the three components accounts for
$\sim 80\%$ of the flux measured at $6\asec$ resolution. The missing flux is recovered
in the $\sim 1\asec$ resolution image and in particular it is associated with the southern component 55b.
We note that the IRAC image of the SMG is not point-like but 
slightly extended and aligned with the two radio components 55a and 55b.
Therefore, we have at least two (55c and the double source 55a+55b), or three (55a, 55b, and 55c), different radio sources associated with different MIR galaxies that will contribute to the measured FIR and sub-millimetre fluxes. The distance between 55a and 55b is $\simeq 1.1\asec$ corresponding
to $\simeq 8.4$ kpc, while component 55c is at $\simeq 18$ kpc, assuming they are both at the same redshift of 55a.
\paragraph{SMM ID 74.} The radio image has a resolution of $3.11\asec \times 1.66\asec$ 
and rms noise of $22\,\mu$Jy\,beam$^{-1}$. This is the faintest radio source among the 12 bright SMGs. At  $6\asec$ resolution is detected with a S/N$=4.3$. The source is heavily resolved when imaged with the ILT, and we had to use a strong tapering on the long baseline in order to obtain a reasonable compromise between sensitivity and angular resolution. The total flux measured at this resolution is consistent, within 10\%, with the peak flux at $6\asec$ resolution. We note that even with the resolution quoted above, the ILT image has a beam area a factor of $\sim $7 lower than that of the $6\asec$ resolution image. The morphology is rather complicated, with a central nuclear component at the position of the MIR galaxy and two
lobes. The south-eastern lobe is more compact (but still resolved out at angular resolution $\lsim 1\asec$) while the north-western lobe is more extended.
Within $\sim 10\asec$ in the north-eastern direction there is another radio source
with a MIR counterpart that may contaminate the measured FIR and sub-millimetre fluxes.
\paragraph{SMM ID 77.} The radio image has a resolution of $0.37\asec \times 0.26\asec$ 
and an rms noise of $18\,\mu$Jy\,beam$^{-1}$. The IRAC image shows two MIR galaxies separated by $2.2\asec$ corresponding to $16$ kpc, assuming they are both at the same redshift. The SMG is identified with the northernmost of the two IRAC objects. At sub-arcsecond resolution, most of the radio emission is associated with the SMG (detected with S/N$\simeq 6$), with very weak emission (S/N$\simeq 3.5$) possibly related to the southern galaxy. In the radio, the SMG shows some extended structure. The recovered radio flux is within $10\%$ of that measured at $6\asec$ resolution.

\section{Discussion}
\label{sec:dis}
\subsection{Multiplicity in bright SMGs}
The high SFRs ($\gsim 10^3$ M$_\odot$yr$^{-1}$) derived from SED fitting for bright SMGs might be overestimated.
Multiple individual objects might contribute to the fluxes derived in the FIR and sub-millimetre, due to the limited angular resolution ($\sim 10\asec$) in these bands. This region of the SED is crucial for deriving the total infrared luminosity and consequently the SFRs. Given the high surface density of deep IRAC observations, as those obtained for the Cosmic Dawn Survey, it is not uncommon to find more than one MIR-emitting object within $\sim 10\asec$ of the SMG. This does not mean that all these foreground or background MIR sources can significantly contribute to the fluxes measured at longer wavelengths. On the other end, if one of these objects also emits in the radio regime, we can rather confidently  assume that it will contribute to the FIR and sub-millimetre flux given the well-known FIR-radio correlation \citep[e.g.][]{1992ARA&A..30..575C}.

We divided the 12 bright SMGs into two main groups.
The first group comprises the five SMGs  without a MIR- and radio-emitting source within $\lsim 10\asec$ (SMM IDs 15, 16, 24, 29, and 77). For some of these sources, there is clearly extended structure in the radio morphology as imaged at sub-arcsecond resolution, and the radio emission can originate from different regions in a single object or different objects. The limited resolution of the MIR mosaic prevents us from classifying these objects as multiple sources.
The second group comprises the SMGs
that show a second MIR- and radio-emitting source within $\sim 10\asec$, in some cases on a few-arcsecond scale (SMM IDs 14, 49, and 74) and in others on a $\lsim 1$ arcsecond
 scale (SMM IDs 3, 12, 47, and 55). In the latter case, the ILT observations 
 allow us to separate different radio components associated with different MIR galaxies or to different regions in the same galaxy. In the following we make the assumption that the resolved IRAC emission observed in these four SMGs is due to the blending of merging or interacting galaxies (but see also the discussion in Sect.~\ref{sec:sfr}). 

We remind the reader that the sample of 12 SMGs has been drawn only 
selecting objects above a threshold flux density 
at $850\mu$m, with an accurate SED fit and within 1.2 deg of the pointing centre of the LOFAR observations.  
We find that 7 of the 12 galaxies (58\%) show some evidence of multiple components that may significantly contaminate the SFR values derived from SED fitting. In five of the seven SMGs, the radio source at the optical position of the SMG is brighter, at 144 MHz, than the other radio sources within the region considered. In two cases is the opposite:
the brighter radio source is not the one coincident with
the optical position of the SMG. In the assumption
that the radio flux reflects the FIR contribution, for five of the
seven multiple SMGs the overestimation of the SFR (based only on the multiplicity argument) should be less than a factor of two.
Considering the small number statistics, the fraction
of multiple systems is consistent with the values in the range $\sim 30\%- 50\%$ derived from other samples of SMGs 
\citep[e.g.][]{2012A&A...548A...4S, 2012ApJ...761...89B,
2013ApJ...768...91H}. It is worth noting that, since in this study we examined the brightest SMGs in the NEP field, due to confusion we can expect an higher 
fraction of multiple sources than that derived from samples selected at lower fluxes.
While the multiple systems on few-arcsecond scales are probably unrelated
objects, the radio components found in the SMM IDs 3, 12, 47, and 55 might be gravitationally bounded. It is well known that SMGs tend to be  strongly clustered \citep[e.g.][]{2001MNRAS.325.1553M, 2004ApJ...611..725B,2012MNRAS.421..284H,2020ApJ...904....2G}, but other studies have suggested that multiple SMGs
might entirely be explained by the low resolution of single-dish surveys 
\citep{2011ApJ...733...92W, 2018MNRAS.476.2278H}. 
Higher resolution NIR imaging and spectroscopic observations are needed to answer the question of whether SMGs 3, 12, 47, and 55 are merging systems at high redshifts or not. In the meanwhile, for the following analysis, we assume that these four SMGs host physically related multiple galaxies and that the host galaxies of each system are at the same redshift derived from the SED fitting. 
With these assumptions, the projected separations of the different galaxies in SMGs 3, 12, 47, and 55 are in the range $\sim$ 5-20 kpc.

\subsection{The AGN contribution in bright SMGs}
\label{sec:AGN}
Shin et al. (2022) derived the AGN fraction contribution, $f_{\rm AGN}$, from the CIGALE-SED fitting \citep{2009A&A...507.1793N, 2019A&A...622A.103B}. The $f_{\rm AGN}$ parameter is defined as the ratio of the AGN luminosity to the  sum of the AGN and dust luminosity.
There is no clear threshold value for the $f_{\rm AGN}$ parameter above which the galaxy emission is dominated by an AGN rather than star formation. Values in the range  $0.1-0.3$ have been used in the past \citep[e.g.][]{2010A&A...514A..10S},
and this combined with the large uncertainties associated with the derived values of $f_{\rm AGN}$ means that such an estimate must be only used as a qualitatively prior for the presence and impact of an
AGN in the SED. For these reasons, based on the $f_{\rm AGN}$ derived by Shin et al.
(2022), we classified the SMGs into three categories: those with  $f_{\rm AGN}>0.3$, which have SEDs clearly dominated by an AGN and which we identify as AGNs in Table~\ref{tab:submm}; those with  $f_{\rm AGN}<0.1,$ which are classified as star-forming galaxies (SFGs) in Table~\ref{tab:submm}; and those with  $0.1\le f_{\rm AGN}\le 0.3,$ which are classified as mixed objects and identified as AGN--SFG in Table \ref{tab:submm}. Half of the SMGs have a derived AGN fraction contribution that places them in the grey zone of AGN--star formation coexistence, only one galaxy is clearly identified as AGN-dominated from the SED fitting and for the remaining 5 galaxies the AGN fraction contribution (again as derived from the SED fitting) is negligible. 
It is worth noting that SED fitting can only take into account 
those AGNs whose torus emission can be traced in the MIR band, i.e. the radiatively efficient AGNs, while
radiatively inefficient AGNs are missed and are usually identified only in the radio domain. In particular,
high-resolution radio observations are necessary
to separate AGN compact radio cores  from more diffuse
radio emission that can be associated with star formation processes. Unfortunately, detecting a compact radio component associated with a SMG does not necessarily mean that the radio emission is AGN-driven as, at the median redshift ($z\sim 2.9$) of our sources, an angular sizes of $\sim 0.3\asec$ corresponds to $\sim 2$ kpc, which is roughly a factor of ten larger than the typical sizes of compact starbursts detected in nearby Ultra Luminous Infrared Galaxies, the local counterparts of high-redshift SMGs \citep[e.g.][]{2005MNRAS.361..748B, 2016A&A...593A..86V, 
2022A&A...664A..25K, 2022A&A...658A...4R}. 
Another way to identify possible AGN radio cores is to derive the brightness temperature ($T_b$) of the compact radio components, since $T_b$ higher than a certain value (depending on the  observing frequency, non-thermal radio spectral index, redshift and the temperature of the gas) can be interpreted as signature of radio emission driven by an 
AGN rather than star formation. This method has been used in the past, in particular using Very Long Baseline Interferometry (VLBI) observations, to discriminate between AGNs and SFGs \citep[e.g.][]{1981ARA&A..19..373K, 1992ARA&A..30..575C}, and more recently using ILT observations \citep{2022MNRAS.515.5758M, 2023A&A...671A..85S}. We calculated the $T_b$ values using the expression

\begin{equation}
T_b= 1.22\times 10^6 \times (1+z) \times \left(\frac{\nu}{1\,{\rm GHz}}\right)^{-2} \times \left(\frac{\theta_{\rm maj}\theta_{\rm min}}{1\,{\rm arcsec}^2}\right)^{-1} \times \frac{S_T} {1\,{\rm Jy}}\, {\rm K,}
\end{equation}
where $z$ is the redshift (given in Table~\ref{tab:submm}), $\nu$ is the observing frequency of 144 MHz, and $S_T$, $\theta_{\rm maj}$ and $\theta_{\rm min}$
are, respectively, the total flux density, and the deconvolved major and minor axis 
of the fitted elliptical Gaussian components (given in Table~\ref{tab:radio}). 
The maximum possible value $T_b^\mathrm{SF}$ for any starburst galaxy is given in \citet{1991ApJ...378...65C} as
\begin{equation}
 T_b^\mathrm{SF}\le T_e\times\left[1+10\times\left(\frac{\nu}{1\,{\rm GHz}}\right)^{0.1-\alpha}\right],
\end{equation}
where $T_e\sim 10^4$ K is the thermal electron temperature and $\alpha$ the radio spectral index following the convention $S\propto\nu^{-\alpha}$. For $\alpha=0.8$ the 
maximum brightness temperature is $\log(T_b^\mathrm{SF})=5.6$ at 144 MHz, and decreases for a flatter spectral index. The brightness temperature values derived for the radio components associated with the SMGs are listed in Table~\ref{tab:radio}.
Of the 12 SMGs, 7 have a radio counterpart with $\log(T_b)>5.6$. Five of these are classified as AGNs or AGNs--SFGs on the basis of the $f_{\rm AGN}$ parameter derived from the SED fitting. The remaining 2 radio sources with a brightness temperature not compatible with radio emission solely due to star formation are classified as pure SFGs from the SED fitting.
These two sources have the highest radio flux densities among the 12 SMGs.

\subsection{Star formation rates from radio emission}
\label{sec:sfr}
All 12 SMGs show either a MIR- and radio-emitting `companion' within $10\asec$ that could contaminate the FIR and sub-millimetre measurements, or a high-brightness temperature radio component, signature of the presence of an AGN. Two sources present both options. Therefore, SFRs derived from the SED fitting could be significantly biased. 
In this section we derive the SFRs from the observed radio emission and compare these values with those obtained from the SED fitting. 

To derive reliable SFRs from the detected radio emission, we
have to ensure that the radio flux we convert is the result of star formation processes and not produced by a radio-emitting AGN. 
In the previous section we have identified the sources where the presence of a radio-emitting AGN is inferred on the basis of the derived brightness temperature.
In this section, for each SMG, we estimate the radio emission that is produced by star formation processes and we convert this to radio luminosity at 144 MHz (L$_{\rm 144MHz}$) and SFR.

To do this, we started from the total flux density measured for each SMG at $6\asec$ resolution. For the sources for which we do not find multiple radio sources on 
sub-arcsecond scales (SMM IDs 14, 15, 16 24, 29, 49, 74, and 77), the radio flux associated 
with star formation is the total flux density measured at $6\asec$ resolution, $S_{\rm SF}=S_{6\asec}$, for the sources with $\log(T_b)<5.6$, or the value obtained subtracting the flux density of the sub-arcsecond component,  $S_{\rm SF}=S_{6\asec}-S_{\rm ILT}$, for the sources with $\log(T_b)>5.6$. In doing this we assume that all the radio emission from a high-brightness temperature component is originated from the AGN and all the remaining radio flux density is produced by
star formation. This might not be entirely true.
The estimate of $T_b$ is sensitive to the angular resolution of the radio image and it is possible that higher resolution images could still detect high $T_b$
components, but with a lower total flux density. Conversely, it is also possible that a fraction of the extended radio emission that we associate with star formation might be originated by the AGN (e.g. 
small jets or lobes).

For each source where we find multiple objects on
sub-arcsecond scales (SMM IDs 3, 12, 47, and 55), 
first we checked whether the flux measured at low resolution, S$_{6\asec}$, is consistent (within 10\%) with the sum of the fluxes of the individual components in the ILT image. If not, the exceeding flux is equally divided among the multiple components (two each for SMM IDs 3, 12, and 47 and three for SMM ID 55). For each component of these four SMGs with 
$\log(T_b)<5.6$ , the radio flux used to derive the SFR is the sum of the flux measured from the ILT image (listed in Table~\ref{tab:radio}),
and the quota of extended flux not recovered in the ILT images and previously derived.
For components with $\log(T_b)>5.6$ only the quota of extended flux is considered. 

\begin{figure}[htp]
 \centering
 \includegraphics[width=8cm]{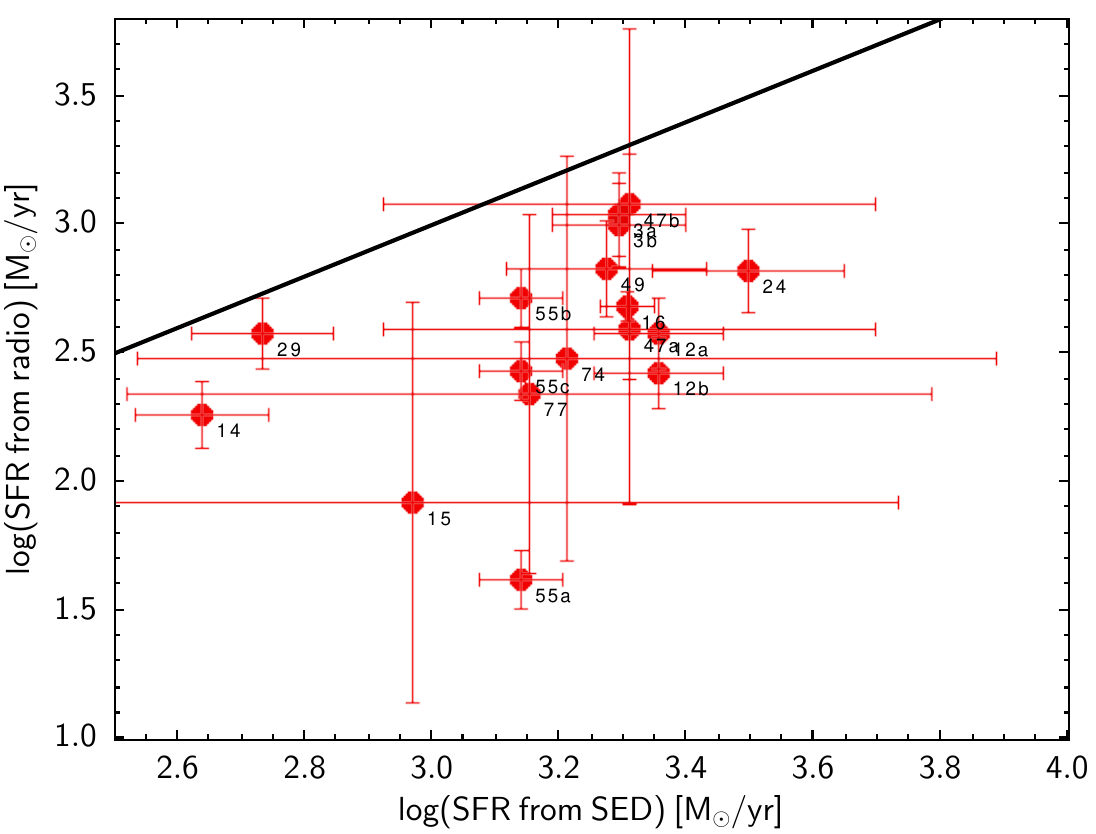}
 \caption{SFR derived using the radio flux versus the SFR derived from SED fitting. For SMGs with multiple radio components, the radio-derived SFR values are plotted for the individual components. The black line is the 1:1 line where the SFRs derived from the two methods are equal.}
 \label{fig:sfr_comp}
\end{figure}

\begin{figure}[htp]
 \centering
 \includegraphics[width=8cm]{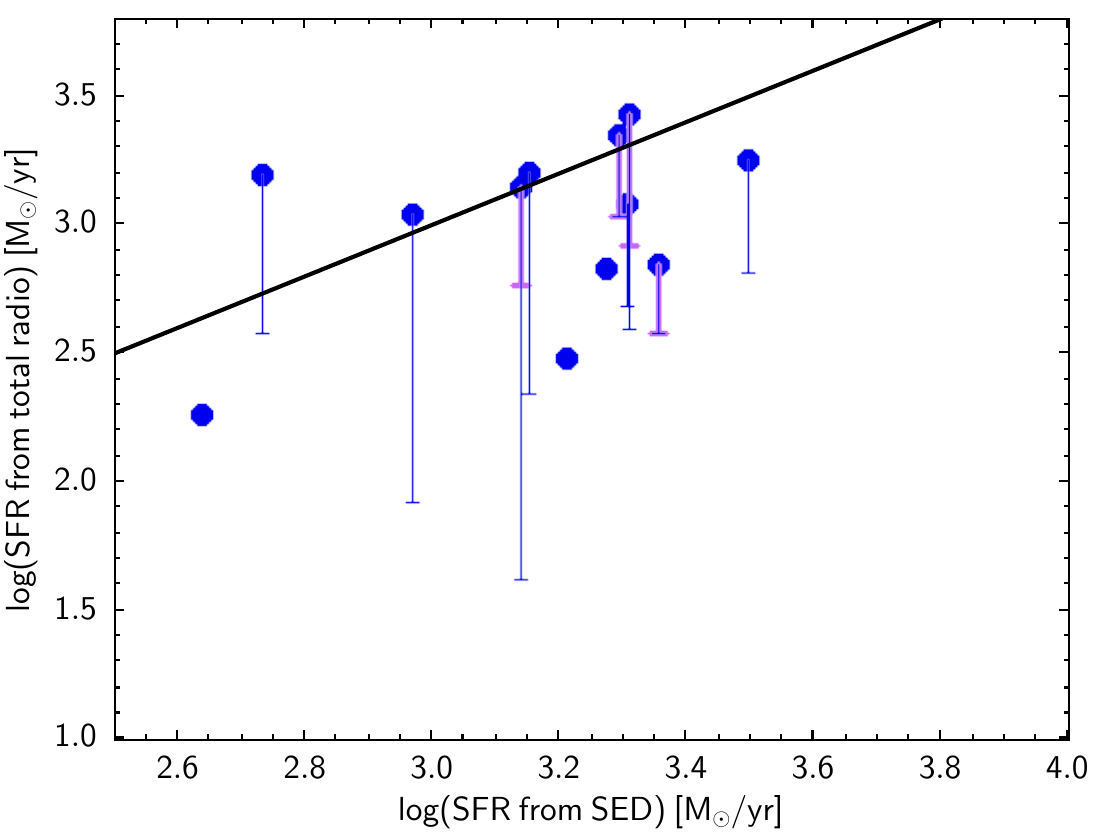}
 \caption{SFR derived using the total radio flux measured at $6\asec$ resolution versus the SFR derived from SED fitting. The blue points have been derived using Eq.3. The purple error bars show the SFRs when only the contribution from radio components on scales $<6\asec$ is subtracted (SMM IDs 3a, 12a, 47a, and 55a), and the blue error bars show the SFRs when the radio AGN contribution is subtracted as well (SMM IDs 15, 16, 24, 29, 47a, 55a, and 77). For SMGs 47a and 55a, the blue error bars take the sum of the two contributions into account. 
 The black line is the 1:1 line where the SFRs derived from the two methods are equal.}
 \label{fig:sfr_tot}
\end{figure}

Having estimated the radio flux density associated with star formation, we followed \citet{2022A&A...658A...1M} to  convert it into a SFR, by using the best-fit parameters for Eq. (2) of \citet{2021A&A...648A...6S}:
\begin{equation}
 \log(\rm SFR)=\frac{
 \log({\rm L_{SF}}) - 22.22 - 0.33\times[log({\rm M_\star}) - 10]}{0.9}, 
\end{equation}
where the monochromatic radio luminosity at 144 MHz is in W\,Hz$^{-1}$, the stellar mass in M$_\odot$ and the derived SFR is in M$_\odot$\,yr$^{-1}$.
The comparison between the SFR values derived from SED fitting and radio emission is shown in Fig.~\ref{fig:sfr_comp}. 
The error bars in this plot are mostly dominated by the uncertainties in the derived photometric redshifts, which are very large for SMM IDs 15, 47, 74, and 77.
The SFR values derived from the radio emission are
lower than those derived from SED fitting by a factor $\sim 1.5$ up to factors $> 10$, with a median ratio of $\sim 5$. 
To further investigate this effect we derived the SFRs using the total radio flux measured at $6\asec$
resolution. These SFR values should be regarded as
upper limits, since they do not take the AGN contribution into account (i.e. the radio flux of the high $T_b$
component is not subtracted), nor the multiplicity on scales $<6\asec$. The results are shown in Fig.~\ref{fig:sfr_tot} using blue dots. For half of the SMGs (SMM IDs 12, 14, 16, 24, 49, and 74)
the SFRs derived from SED fitting are still larger (by factors in the range $1.7-5.4$) than those obtained from the total radio flux. Among these six sources there are the three SMGs (SMM IDs 14, 49, and 74) without a high $T_b$ component and with a possible FIR confusing source at distances $\sim 10\asec$. Five SMGs show SFR values 
consistent between the two methods: these SMGs are three of the four SMGs with multiplicity on scales $<6\asec$ (SMM IDs 3, 47, and 55) and two SMGs with high $T_b$ (SMM IDs 15 and 77). 
The only SMG with a SFR, derived from the total radio flux, significantly higher than the SFR derived from SED fitting is SMM ID 29, which is unambiguously classified as AGN-dominated 
by the SED fitting in \citet{2022MNRAS.514.2915S} and by the radio properties as well. 
The purple error bars show the SFRs that we obtain once the multiple radio components on scales $<6\asec$ are subtracted. This effect accounts for a decrease of $\sim 0.3-0.5$ dex in the SFR values derived from radio,
and affects mostly three SMGs (SMM IDs 3a, 47a, and 55a).
The blue error bars show the SFRs that we obtain 
once the radio AGN contribution is subtracted as well.
Only this effect counts for a decrease of $\sim 0.3-1.0$ dex in the SFR values derived from radio 
(e.g.  SMM IDs 15, 29, 55a, and 77).
Summarising, correcting for the multiplicity and the presence of a radio AGN clearly affects the SFR values derived from the radio emission, but even using the total radio flux measured at $6\asec$ resolution, without taking into account the multiplicity and AGN contribution, we find that for half of the SMGs the SFRs derived from SED fitting are larger than those derived from radio emission.

\begin{figure}[htp]
 \centering
 \includegraphics[width=8cm]{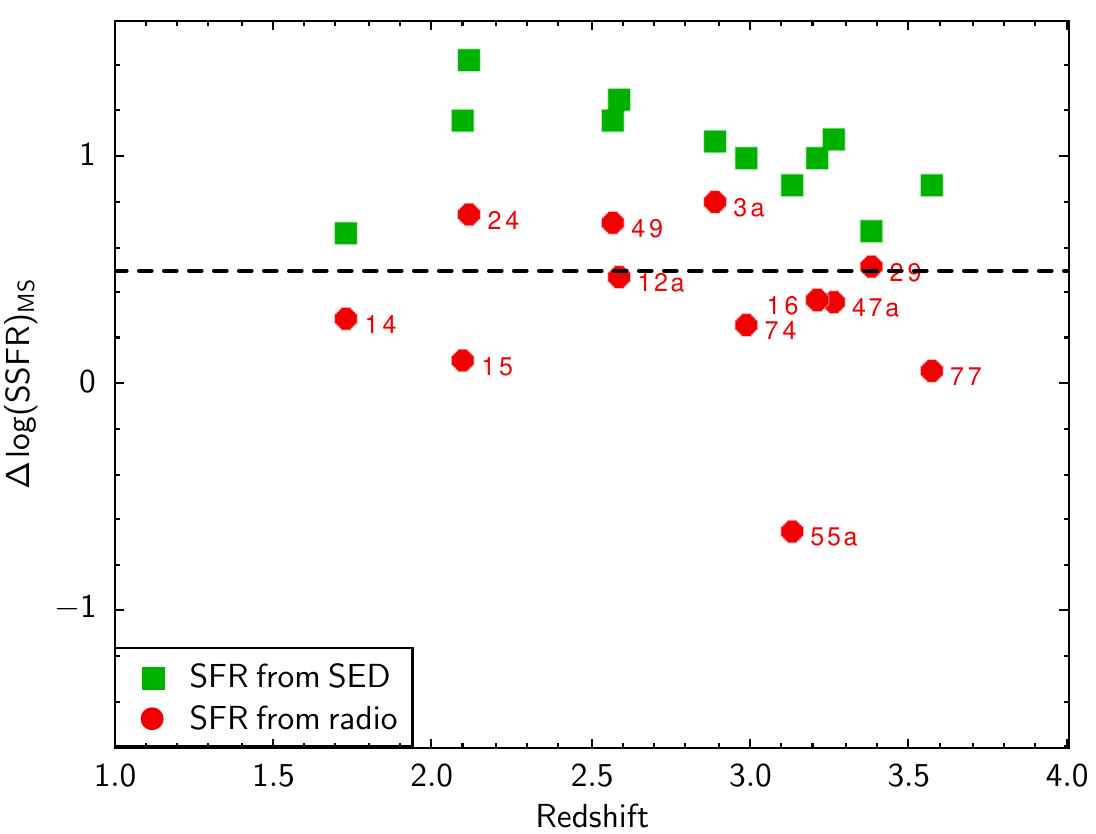}
 \caption{Offset of the specific SFR from the SFG MS derived using the fit in \citet{2023MNRAS.519.1526P} plotted as a function of redshift. Green squares are $\Delta\log(\mathrm{SSFR})_\mathrm{MS}$ values derived using the SFRs from SED fitting, and red dots those obtained using SFRs from the radio. Red dots are labelled with the SMM ID of the SMG. For SMGs with multiple components, we only show the $\Delta\log(\mathrm{SSFR})_\mathrm{MS}$ of the component associated with the SMGs. The dashed black line is drawn at $\Delta\log(\mathrm{SSFR})_\mathrm{MS}=0.5$; galaxies above this line have three times the SFR of a MS galaxy with the same stellar mass and redshift.}
 \label{fig:sfr_ms}
\end{figure}

In Fig.\,\ref{fig:sfr_ms} we plot the distance of our SMGs from the SFG MS in the SFR-stellar mass plane
defined as
\begin{equation}
\Delta\log(\mathrm{SSFR})_\mathrm{MS}=\log(\mathrm{SSFR}) -\log[\mathrm{SSFR}_\mathrm{MS}(M_\mathrm{star,z})].
 \end{equation}
  To parameterise the MS of SFG as a function of redshift and stellar mass, we used the best-fit
relation from \citet{2023MNRAS.519.1526P}.  \citet{2023MNRAS.519.1526P} extended the analysis carried out in \citet{2014ApJS..214...15S} by including more recent studies, which allowed them to probe the evolution of the MS of SFGs over the widest range of redshift ($0<z<6$) and stellar mass ($10^{8.5} - 10^{11.5}$ M$_\odot$).
This can be relevant for the present discussion, since SMGs are usually hosted by the most massive galaxies. 

The advantage of using $\Delta\log(\mathrm{SSFR})_\mathrm{MS}$ is to remove the effects of different stellar mass and redshift evolution and to easily distinguish
galaxies with SFRs typical of MS objects ($\Delta\log(\mathrm{SSFR})_\mathrm{MS}\sim 0$) from strong starburst galaxies that usually are defined as those with a SFR larger than 3 times that of a MS galaxy ($\Delta\log(\mathrm{SSFR})_\mathrm{MS}\gsim 0.5$).
 Figure\,\ref{fig:sfr_ms} shows both the offset with respect to the SFG MS obtained using the SFR from SED fitting (green squares) derived in \citet{2022MNRAS.514.2915S} and that derived using the SFR from radio (red points). Using the whole sample of SMGs in the NEP region (not just the brightest ones) and the SFRs derived from the SED fitting, \citet{2022MNRAS.514.2915S} find that $\sim 40\%$ of SMGs are strong starburst galaxies with
 $\Delta\log(\mathrm{SSFR})_\mathrm{MS}> 0.5$, slightly decreasing for galaxies at $z\gsim 3$. 
Using the SFR values derived from SED fitting, all 12 bright SMGs have 
$\Delta\log(\mathrm{SSFR})_\mathrm{MS}\gsim 0.5$, with a median value and scaled median absolute deviations of $1.02\pm 0.21$.
It is worth noting that \citet{2022MNRAS.514.2915S} used the 
\citet{2014ApJS..214...15S} best-fit that does not include the curvature of the MS relation at high stellar masses and high redshifts found by \citet{2023MNRAS.519.1526P}. Using the relation derived by \citet{2023MNRAS.519.1526P} yields to larger differences between the measured specific SFR and those expected from MS SFGs with respect to the values presented in \citet{2014ApJS..214...15S}.
Using the radio flux as a proxy for SFR,  
we find that only about one-third of the bright SMGs are strong starburst galaxies with specific SFR values larger than three times that of a MS SFG, and for only one object is the SFR less than that expected from a MS SFG with the given mass and redshift. 
The overall median value and scaled median absolute deviation of the red points plotted in Fig.\,\ref{fig:sfr_ms}
is $0.36\pm 0.30$.
The sample of 12 SMGs is too small to see a significant trend of $\Delta\log(\mathrm{SSFR})_\mathrm{MS}$
with redshift as noted by \citet{2014ApJS..214...15S}. The only speculative interpretation is that the higher-redshift 
($z\gsim 3$) bright SMGs in Fig.\,\ref{fig:sfr_ms} are associated with host galaxies whose SED is best fitted by a hybrid SFG--AGN component (see Table~\ref{tab:submm}) and  have slightly lower specific SFR values than lower-redshift ($z<3$) bright SMGs, which are mostly associated with SFGs.

\begin{figure}[htp]
 \centering
 \includegraphics[width=8cm]{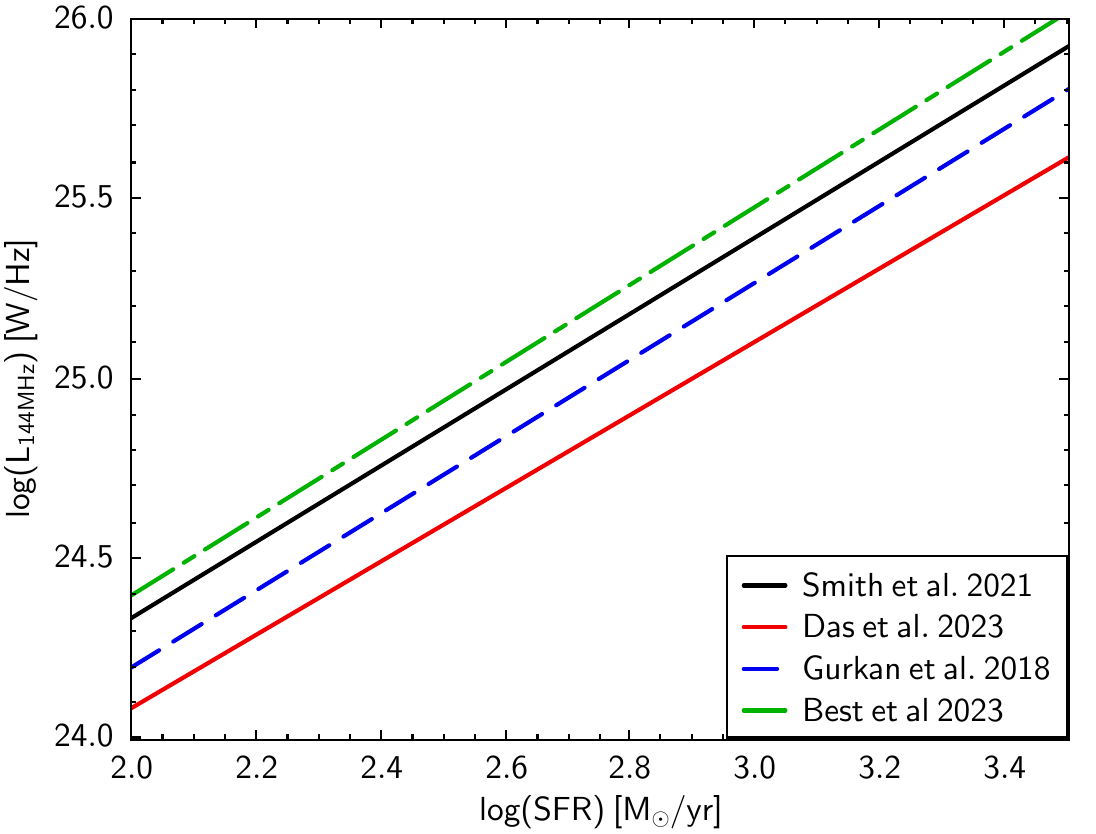}
 \caption{Comparison of four L$_{\rm 144MHz}$-SFR calibrations derived from LOFAR observations at 144 MHz. For an easier comparison we use the calibrations obtained without the stellar mass dependence: 
 $\log(L_{\rm 144MHz})=22.221+1.058\log({\rm SFR})$ \citep{2021A&A...648A...6S},
 $\log(L_{\rm 144MHz})=22.024+1.019\log({\rm SFR})$ \citep{2024MNRAS.531..977D}, 
 $\log(L_{\rm 144MHz})=22.06+1.017\log({\rm SFR})$ \citep{2018MNRAS.475.3010G},
and  $\log(L_{\rm 144MHz})=22.24+1.08\log({\rm SFR})$ \citep{2023MNRAS.523.1729B}.
 To better highlight the differences between calibrations, the plot shows only the region with high 
 L$_{\rm 144MHz}$ and SFR. When needed, the SFR was converted to the initial mass function of 
 \citet{2003ApJ...586L.133C} using the corrections provided in \citet{2014ARA&A..52..415M}.}
 \label{fig:lum_sfr_cal}
\end{figure}

\subsection{Caveats: How our assumptions affect the SFR comparison}
Several assumptions made in the previous analysis can significantly affect the comparison between the SFR derived from SED fitting and that obtained from L$_{\rm 144MHz}$. Here we discuss the most relevant ones.

The first point we want to address is the calibration of the L$_{\rm 144MHz}$-SFR relation. We used the mass-dependent L$_{\rm 144MHz}$-SFR calibration derived by \citet{2021A&A...648A...6S} and given in Eq.
3. The main reasons why we adopted this calibration are: (1) it is based on a large ($\sim 118,000$) NIR selected sample of galaxies, which allows us to properly take the galaxies not detected in the radio into account and to avoid the selection effects introduced by radio-selected or FIR selected samples; (2)
the calibration has been obtained at 144 MHz, the same frequency of our observations, meaning it is not necessary to add a further assumption on the radio spectral index;
and (3) the SFRs are derived using the SED fitting code \texttt{\small MAGPHYS}, which has been extensively tested and widely used; this enables an easier comparison with other calibrations.
As reported by  \citet{2021A&A...648A...6S} the scatter
observed in the L$_{\rm 144MHz}$--SFR relation is 
of about 0.3 dex around SFR $\sim 100$ and possibly increasing at higher SFRs. Other studies found a similar scatter \citep[e.g.][]{2017MNRAS.469.3468C, 2018MNRAS.475.3010G, 2023MNRAS.523.1729B, 2023MNRAS.523.6082C, 2024MNRAS.531..977D}. In Fig.~\ref{fig:lum_sfr_cal} we show the L$_{\rm 144MHz}$-SFR calibrations, limited to the region of
high SFRs,
derived by four studies based on radio observations obtained by LOFAR at 144 MHz. For a more immediate comparison we show the mass-independent calibrations, but we note that the differences
in the SFR values obtained from using a mass-dependent
rather than a mass-independent calibration are small.
For instance, for our 12 SMGs, the distribution of this difference has a median of 0.015 dex and a scaled MAD of 0.125 dex, much smaller than the typical dispersion.
The plot clearly highlights that the four calibrations have systematic offsets but they are usually smaller than 0.3 dex, and in particular the calibration of \citet{2021A&A...648A...6S} is intermediate between the relations derived by \citet{2023MNRAS.523.1729B} and  \citet{2018MNRAS.475.3010G}.
The calibration obtained by 
\citet{2024MNRAS.531..977D} uses the rather new SED fitting code \texttt{\small PROSPECTOR} and it is possible that there could be systematic differences between the SFRs derived by \texttt{\small MAGPHYS}
and \texttt{\small PROSPECTOR}. It is worth noting that studies based on radio observations at 1.4 GHz find similar differences in the L$_{\rm 144MHz}$-SFR calibration \citep[see Appendix C in][for a detailed discussion]{2014ApJS..214...15S}.

The second assumption that can have a significant impact on the SFRs derived from the measured L$_{\rm 144MHz}$ is the contribution of the radio AGN.
It is possible that we are overestimating the number
of SMGs with a radio AGN and/or we are overestimating the radio flux originated by the AGN that we subtracted to obtain the radio flux from star formation processes. As can be seen in Fig.~\ref{fig:sfr_tot} subtracting the radio flux 
assumed to be originated by the AGN can have a strong effect on the derived SFRs. Two points need to be considered.
Firstly, the number of SMGs with a radio AGN depends on the adopted threshold value in $T_b$. We derived the threshold $T_b^{\rm SF}$ at 144 MHz assuming 
$\alpha=0.8$ and $T_e=10^4$\,K. The assumption on the spectral index is conservative: we assumed a value that is typically observed in SFGs
at $\sim 1$ GHz \citep[e.g.][]{2015A&A...573A..45M},
while studies exploring the radio spectral properties of SFGs at lower frequencies find
flatter spectral index $\alpha\simeq 0.5$ possibly due to thermal free-free absorption \citep[e.g.][]{2024MNRAS.528.5346A}. A radio spectral index flatter than 0.8 would yield to a lower value for $T_b^{\rm SF}$ and therefore to more SMGs with a radio AGN core. On the other end, such an effect could be compensated for by a larger value of the thermal electron temperature with respect to the assumed value, even if there is no clear evidence of this in the literature and the value $T_e=10^4$ K has been 
generally used for compact starburst \citep{1991ApJ...378...65C} and Arp 220, an ultra-luminous infrared galaxies, \citep{2000ApJ...537..613A}.
Secondly, it is possible that by subtracting the flux density of the high $T_b$ component we are also subtracting radio emission that is not produced by the AGN.
This cannot be excluded, but, to be fair, we cannot exclude the opposite, i.e. that at least part of the remaining radio emission used to calculate the SFR is not produced by star formation but is extended emission generated by the AGN via short jets or outflows.
Summarising, the assumptions adopted to derive the AGN contribution to the measured radio flux densities could not always be correct, to some extent, for all the SMGs, but at least
none of them should produce the effect of systematically underestimate the SFRs derived from L$_{\rm 144MHz}$.
In any case, we note that most of the SMGs have $T_b$ that are within $\pm 0.2$ dex of the threshold value and therefore even modest changes in our assumptions could have a significant impact in increasing or decreasing the derived $T_b$.

Finally, we assumed that the different radio components detected in the range 5-20 kpc in four SMGs
and visually coincident to other regions of IRAC emission are different galaxies, maybe interacting and physically bound or just close in projection, and therefore should be considered separately. The alternative interpretation would be a single galaxy with regions of bursting star formation separated by several kiloparsecs that are also bright at $4.5\mu m$. 
We believe the first hypothesis is more realistic and that these are typical examples of SMGs where the extremely high SFRs are the result of a FIR and sub-millimetre photometry that cannot separate two or
more, possibly merging, galaxies emitting in these bands.

It is important to note that even dismissing
the last two assumptions and using the L$_{\rm 144MHz}$ derived from the total $6\asec$ resolution image, the radio-derived SFR for half of the SMGs is already lower than that obtained from SED fitting.
To summarise, while the assumptions we have adopted can easily account for a $\sim 0.3-0.5$ dex in the dispersion of the radio-derived SFRs, we cannot conclude that such a dispersion would only produce lower, rather than also higher, SFRs.

\section{Summary}

We have presented the results obtained imaging a sample of bright
SMGs in the NEP region with the ILT at sub-arcsecond angular resolution. The SMGs were selected from the catalogue published by \citet{2022MNRAS.514.2915S} on the basis of the following criteria: brighter than 9 mJy at $850\mu m$, classified as having a reliable SED fit, and within 1.2 deg of the pointing centre of the ILT observations. Using these criteria, we selected 12 bright SMGs. All these SMGs have extreme SFR values, derived from SED fitting, in the range $\sim 500-3000$ M$_\odot$yr$^{-1}$ \citep{2022MNRAS.514.2915S}. 

Our main results can be summarised as follows:
\begin{enumerate}
\item All 12 SMGs are detected in the sub-arcsecond LOFAR images
with $S/N > 5$, with the exception of one SMG that has $S/N \sim 4$.

\item Using our LOFAR observations and the deep IRAC images publicly available in
  this field, we find evidence of possible multiplicity in 7 out of the 12 SMGs, meaning that we have identified another radio- and MIR-emitting object within $\sim 10\asec$ of the SMG. The presence of
  such objects could contaminate the FIR and sub-millimetre fluxes used in the SED fitting and consequently lead to overestimate the derived SFRs.
  For four out of the seven SMGs, the multiple galaxies are found on sub-arcsecond scales,
  corresponding to projected separations in the range $\sim 5-20$ kpc,
  assuming the multiple objects are at the same redshift as the SMG.

  \item Using the sub-arcsecond-scale ILT observations, we derived the $T_b$ of the compact radio components. Out of the 12 SMGs, 7 have 
  a compact radio component with $T_b$ exceeding the maximum $T_b$ compatible with radio emission produced by star formation processes. In these objects, the compact radio cores must be associated with a radio-emitting AGN.
  Again, the presence of the AGN might have significantly contributed to
  the extremely high SFRs derived by the SED fitting.
  We note that all 12 SMGs are affected by multiplicity and/or the
  presence of a radio-emitting AGN.
  
  \item To derive an alternative estimate of the SFR in the 12 bright SMGs, one that is independent of the SED, we combined the information obtained from
  the ILT observations at $6\asec$ and at sub-arcsecond resolution
  to disentangle the contribution of the  AGNs from the total radio emission of the SMGs and their possible companions. Assuming that the radio emission not directly associated with the AGN is due to star formation, we used this information to derive the SFR using the 
  relation found by \citet{2021A&A...648A...6S}. The SFRs derived in this way are lower than those derived from SED fitting, on average by a factor of $\sim 5$.
  
  \item 
  We compared the offset of the specific SFR from the SFG MS ($\Delta(\mathrm{SSFR})_\mathrm{MS}$) using both the SFRs derived from the SED fitting and those derived from the radio emission. The median values are $1.02\pm 0.21$ and
  $0.36\pm 0.30$, respectively, and the errors are the scaled median absolute deviations. 
  Using the radio emission as a proxy for the determination of the SFRs shows that bright SMGs are hosted by SFGs or hybrid SFG--AGN systems, which, on average, are only a factor of 2 more star-forming than the MS objects.
  
  \item
  Finally, we have discussed how the assumptions adopted
  in our analysis can affect the results and in particular the radio-derived SFR values. The L$_{\rm 144MHz}$--SFR calibration and the identification and subtraction of a radio AGN component are the more impactful assumptions, and they can easily introduce a scatter of $\sim 0.3-0.5$ dex in the radio-derived SFRs. This means that by using different choices  we could have obtained quantitatively slightly different results, but we argue that none of these assumptions is extreme or necessarily introduces a systematic offset, and therefore, in principle, we could have obtained higher or lower radio-derived SFRs.
\end{enumerate}
\begin{acknowledgements}
We thanks the anonymous referee for very helpful and constructive comments.
Marco Bondi, IP, MM, LB, MB, MG, RS  acknowledge support from INAF under the Large Grant 2022 funding scheme (project "MeerKAT and LOFAR Team up: a Unique Radio Window on Galaxy/AGN co-Evolution”. This paper is based on data obtained with the International LOFAR Telescope (ILT) under project codes LC12\_027.

LOFAR is the Low Frequency Array, designed and constructed by ASTRON. It has observing, data processing, and data storage
facilities in several countries, which are owned by various parties (each with their own funding sources), and which are collectively operated by the ILT foundation under a joint scientific policy. The ILT resources have benefited from the following recent major funding sources: CNRS-INSU, Observatoire de Paris and Université d’Orléans, France; BMBF, MIWF-NRW, MPG, Germany; Science Foundation Ireland (SFI), Department of Business, Enterprise and Innovation (DBEI), Ireland; NWO, The Netherlands; The Science and Technology Facilities Council, UK; Ministry of Science and Higher Education, Poland; The Istituto Nazionale di Astrofisica (INAF), Italy.

This research made use of the OCCAM supercomputing facility 
run by the Competence Centre for Scientific Computing an interdepartmental advanced research centre, that specializes in High Performance Computing (HPC).

This research made use of the LOFAR-IT computing infrastructure supported and operated by INAF, including the resources within the PLEIADI special "LOFAR" project by USC-C of INAF, and by the Physics Dept. of Turin University (under the agreement with Consorzio Interuniversitario per la Fisica Spaziale) at the C3S Supercomputing Centre, Italy.
\end{acknowledgements}

\bibliographystyle{aa} 
\bibliography{paper-scuba}

\end{document}